# On-chip multi-color microdisk laser on Yb$^{3+}$-doped thin-film lithium niobate

YUAN ZHOU,[1,3] ZHE WANG,[1,3,4] ZHIWEI FANG,[2,8] ZHAOXIANG LIU,[2,9] HAISU ZHANG,[2] DIFENG YIN,[1,3] YOUTING LIANG,[2] ZHIHAO ZHANG,[1,3,4] JIAN LIU,[2] TING HUANG,[2] RUI BAO,[2] RONGBO WU,[1,3] JINTIAN LIN,[1] MIN WANG,[2] AND YA CHENG[1,2,5,6,7,10]

[1]*State Key Laboratory of High Field Laser Physics and CAS Center for Excellence in Ultra-intense Laser Science, Shanghai Institute of Optics and Fine Mechanics (SIOM), Chinese Academy of Sciences (CAS), Shanghai 201800, China*
[2]*The Extreme Optoelectromechanics Laboratory (XXL), School of Physics and Electronic Science, East China Normal University, Shanghai 200241, China*
[3]*Center of Materials Science and Optoelectronics Engineering, University of Chinese Academy of Sciences, Beijing 100049, China*
[4]*School of Physical Science and Technology, ShanghaiTech University, Shanghai 200031, China*
[5]*Collaborative Innovation Center of Extreme Optics, Shanxi University, Taiyuan 030006, China.*
[6]*Collaborative Innovation Center of Light Manipulations and Applications, Shandong Normal University, Jinan 250358, People's Republic of China*
[7]*Shanghai Research Center for Quantum Sciences, Shanghai 201315, China*
[8]*zwfang@phy.ecnu.edu.cn*
[9]*zxliu@phy.ecnu.edu.cn*
[10]*ya.cheng@siom.ac.cn*

**We demonstrate an on-chip Yb$^{3+}$-doped lithium niobate (LN) microdisk laser. The intrinsic quality factors of the fabricated Yb$^{3+}$-doped LN microdisk resonator are measured up to 3.79 × 10$^5$ at 976 nm wavelength and 1.1 × 10$^6$ at 1514 nm wavelength. The multi-mode laser emissions are obtained in a band from 1020 nm to 1070 nm pumped by 984 nm laser and with the low threshold of 103 μW, resulting in a slope efficiency of 0.53% at room temperature. Furthermore, the second-harmonic frequency of pump light and the sum-frequency of the pump light and laser emissions are both generated in the on-chip Yb$^{3+}$-doped LN microdisk benefited from the strong $\chi^{(2)}$ nonlinearity of LN. These microdisk lasers are expected to contribute to the high-density integration of LNOI-based photonic chip.**

Featured with the long lifetime, high quantum efficiency, broad gain bandwidth, and easy incorporation, the ytterbium ion (Yb$^{3+}$) has been successfully used as a doping element in laser media, specifically in solid-state lasers and double-clad fiber lasers with high-efficiency, high-power, and mode-locked outputs [1-4]. As a result, the Yb$^{3+}$-doped laser has become the workhorse of high-power laser systems prevalently adopted in industry and defense [5]. The on-chip microlaser offers intriguing properties of ultra-compact, low-cost, and high-performance, so as to be considered as scalable light sources for a broad range of applications including high-speed optical communications, sensing, and quantum technologies [6-8]. The on-chip Yb$^{3+}$-doped microlasers based on microresonators have been previously demonstrated on the silica and aluminum oxide (Al$_2$O$_3$) substrates, presenting great utilities for biosensing applications and integrated photonic chips [9-13]. On the other hand, the lithium niobate (LN) crystal provides an attractive option to be the host material for rare earth ions, owing to their broad optical transparency window (0.35–5 μm), low optical loss (~0.1%/cm @1064 nm), high refractive index (~2.2), high nonlinear coefficient ($d_{33} = -27.2 \pm 2.7\ pm/V@\lambda = 1.064\ \mu m$), and large electro-optical effect ($r_{33} = 30.9 pm/V@\lambda = 632.8 nm$) [14-16]. The on-chip microlasers and amplifiers based on the Er$^{3+}$-doped thin film lithium niobate on insulator (TFLNOI) have been demonstrated recently, showing great promise for high-performance scalable light sources on integrated photonics [17-26].

In this Letter, we demonstrate a monolithic on-chip Yb$^{3+}$-doped LN microdisk laser. Pump light injection and laser emissions out-coupling are achieved with a tapered fiber. We have measured the intrinsic quality (Q) factors up to 3.56 × 10$^5$ at 976 nm wavelength and 1.1 × 10$^6$ at 1514 nm wavelength in the fabricated Yb$^{3+}$-doped LN microdisk. We observe the multi-mode laser emissions in a band from 1020 nm to 1070nm pumped at the wavelength around 984nm. The threshold of the Yb$^{3+}$-doped LN microdisk laser has been measured to be as low as 103 μW with the slope efficiency of 0.53% at room temperature. Thanks to the strong $\chi^{(2)}$ nonlinearity of LN, the second-harmonic frequency of pump light and the sum-frequency of the pump light and laser lines are also generated in the $Yb^{3+}$-doped LN microdisk. This unique property of the $Yb^{3+}$-doped LN microdisk laser has shown great potential in wideband tunable laser emissions from visible to infrared (IR). These microdisk lasers are expected to contribute to the high-density integration of LNOI-based photonic chip.

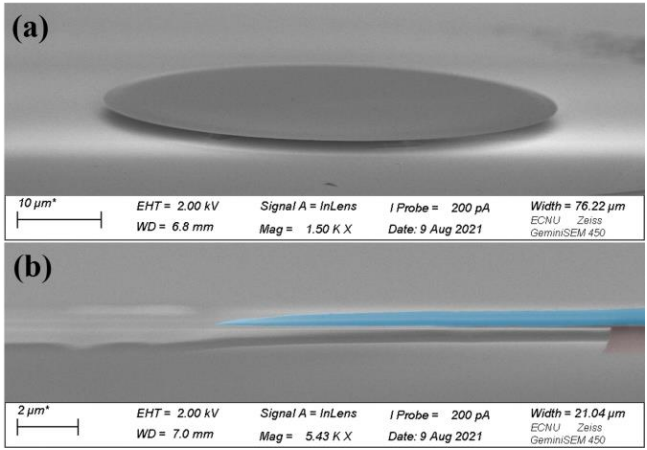

Fig. 1. (a)The scanning electron microscope (SEM) images of the fabricated on-chip $Yb^{3+}$-doped LN microdisk. (b) The side view false color SEM where the $Yb^{3+}$-doped LN microdisk is shown in blue and the silica pillars is shown in brown.

The on-chip $Yb^{3+}$-doped LN microdisk was fabricated on a 600-nm-thickness Z-cut LN thin film at an $Yb^{3+}$-doping concentration of 0.5 mol%. The $Yb^{3+}$-doped LN microdisks are fabricated by photolithography-assisted chemomechanical etching (PLACE), the details about the LN microdisk fabrication can be found in our previous work [27, 28]. Figure 1(a) shows the scanning electron microscope (SEM) image of the fabricated $Yb^{3+}$-doped LN microdisk. The $Yb^{3+}$-doped LN microdisk has a diameter of about 53 μm. Figure 1(b) shows the side view false color SEM where the LN microdisk is shown in blue and the silica pillars is shown in brown. The thickness of the $Yb^{3+}$-doped LN disk is approximately 572 nm and the sidewall angle is approximately 9°. The fabricated LN disk is slightly thinner than the original thickness of the LN thin film due to the second chemomechanical polish step in the microdisk fabrication process. The $Yb^{3+}$-doped LN microdisks are supported by silica pillars.

The experimental setup to characterize the laser emission performance of the $Yb^{3+}$-doped LN microdisk is illustrated in Fig. 2(a). Here, a continuously tunable laser of wavelength setting at around 970 nm and 1550 nm (TLB 6712 and 6728, New Focus Inc.) were used for characterizing the Q factors of the microdisk around and away from the pump laser wavelength., respectively, as the continuously tunable laser of 970 nm wavelength was used to pump the $Yb^{3+}$-doped LN microdisk. A fiber taper with a diameter of around 1 μm was used to couple the pump into the $Yb^{3+}$-doped LN microdisk as well as to collect the generated laser emissions from the LN microdisk. The polarization of the pump light was adjusted by a polarization controller (FPC561, Thorlabs Inc). The spectra of the output beam were measured by an optical spectrum analyzer (OSA: AQ6370D, YOKOGAWA Inc.). In addition, the ultraviolet-visible spectra of the output is analyzed by an ultraviolet-visible spectrometer (NOVA, Shanghai Ideaoptics Corp., Ltd). The power of the input pump laser was monitored by a power meter (PM100D, Thorlabs Inc.). A photodetector (New focus 1811, Newport Inc.) was palced in the fiber path for the Q factor measurements of resonant modes in the microdisk. Figure 2(b) and (c) shows the transmission spectra of the $Yb^{3+}$-doped microdisk at the wavelengths of 975.78 nm and 1514 nm, respectively. The Lorenz fitting (red curves) shows that the $Yb^{3+}$-doped microdisk has intrinsic Q factors of $3.79\times10^5$ and $1.09 \times10^6$ (loaded Q factors of $2.55\times10^5$ and $5.46 \times10^5$) at the wavelengths of 975.78 nm and 1514 nm, respectively. The Q factor around 790 nm wavelength is lower than that around the 1550 nm wavelength because of the higher intrinsic absorption of $Yb^{3+}$ ions around 970 nm wavelength. And the Q factor of laser emissions (1000 nm - 1100 nm) is close to the Q factor of 1510 nm due to their similar absorption coefficients.

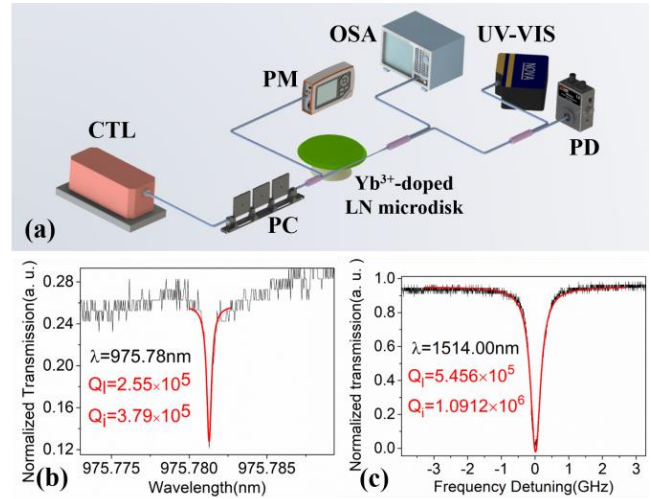

Fig. 2. (a) Experimental setup to characterize the on-chip $Yb^{3+}$-doped LN microdisk laser. (b) and (c) The transmission spectra (black dots) and Lorenz fitting (red curves) show that the $Yb^{3+}$-doped LN microdisk has intrinsic quality factors of $3.79\times10^5$ and $1.09 \times10^6$ at a wavelength of 975.78 nm and 1514 nm, separately. ( CTL: continuously tunable laser; PC: polarization controller; PM: power meter; OSA: optical spectrum analyzer; UV-VIS: ultraviolet-visible spectrometer; PD: photodetector.)

The $Yb^{3+}$-doped LN microdisk laser performance are characterized at a pump wavelength of 984 nm for the sake of the highest laser emissions as observed in the $Yb^{3+}$-doped LN microdisk, and the laser emissions can be expected to span a wavelength range from ∼1020 nm to 1070 nm. The pump and laser wavelengths correspond to the transitions between the ground-state manifold $^2F_{7/2}$ and the excited-state manifold $^2F_{5/2}$ of the ytterbium ion. Figure 3(a) shows the spectra of the $Yb^{3+}$-doped LN microdisk laser at the different pump powers. Figure (b) shows the multi-mode lasing at the wavelengths around 1030 nm and 1060 nm observed under the pump power at 617 μW. The mode wavelength interval is consistent with the free spectrum range (FSR) of the LN microdisk. The dependence of the lasing power of the $Yb^{3+}$-doped LN microdisk laser on the injected pump power is illustrated in Figure 3(c). By linear fitting, the threshold of the lasing mode is found to be around 103 μW, which depends on the coupling condition between the microdisk and fiber taper as well as the thermal condition of the microdisk. Furthermore, the slope efficiency is derived as 0.53 %, which is much higher than the previously demonstrated $Er^{3+}$-doped LN microdisk due to the higher quantum efficiency of the ytterbium ions [17-19]. The higher quantum efficiency also leads to lower pump threshold as shown in Figure 3(c).

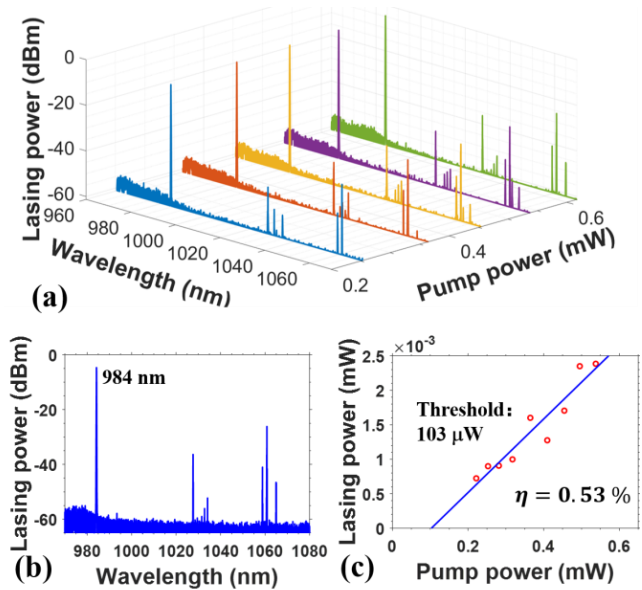

Fig. 3 Lasing characterization of the Yb$^{3+}$-doped LN microdisk under a pump of 984 nm wavelength. (a) Evolution of the lasing modes under different pump power. (b) Multi-mode lasing under the pump power at 617 μW. (c) The Yb$^{3+}$-doped LN microdisk lasing power with the increasing pump power. The blue line is the linear fitting of the lasing power. A lasing threshold can be derived at 103 μW and the slope efficient of the lasing is around 0.53 %.

Thanks to the strong χ$^{(2)}$ nonlinearity of LN, the second harmonic of the pump light and the sum-frequency between the pump laser and the induced laser emissions are also generated directly in the Yb$^{3+}$-doped LN microdisk as shown in Figure 4. When tuning the pump laser wavelength around 969 nm, there are two lasing lines appeared at 1031.5 nm and 1061.8 nm, respectively. Meanwhile, three new emissions at 484.5 nm, 499.7 nm, and 507 nm are also generated from the microdisk as detected by the fiber spectrometer in the visible spectral region, as shown in the left inset in Figure 4. As the arc arrows clearly denoted in Figure 4, the emission line at 484.5 nm is generated through the second harmonic generation (SHG) of the pump light, and the other two lines arise from the sum-frequency generation (SFG) between the pump laser and the generated microdisk laser.

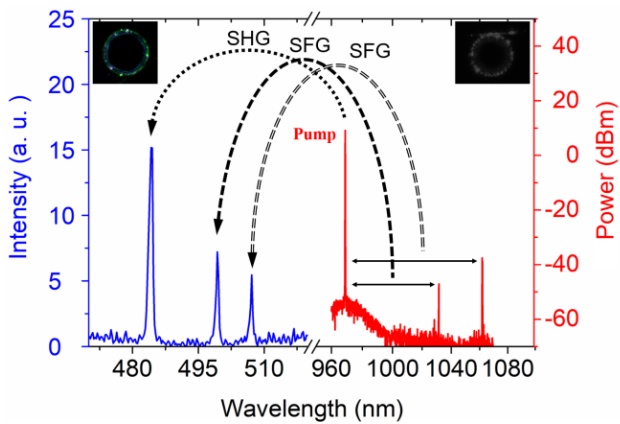

Fig.4. Intra-cavity frequency conversions from the pump laser and the Yb$^{3+}$-doped LN microdisk laser to generate new frequency component at visible wavelength through SHG and SFG. Insets shows the photograph of the microdisk at the visible (left) and infrared (right) regions.

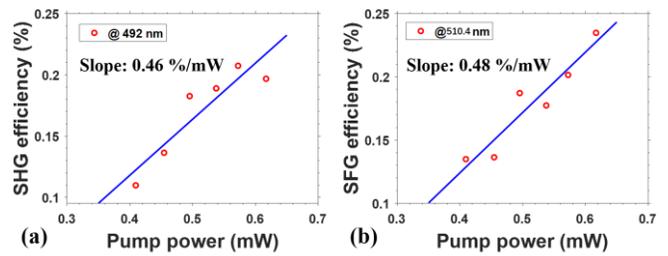

Fig. 5 (a) The SHG and (b) SFG conversion efficiency as a function of the pump power..

Furthermore, by optimizing the pump laser wavelength around 984 nm, a brighter laser emission than that observed with the pump wavelength around 969 nm and a stronger second-order nonlinear effect have been observed. We observe an emission line at 492 nm which can be attributed to the SHG of the pump light, and another emission line at 510.4 nm originating from the SFG between the pump laser at 984 nm and the laser line at 1060 nm. It was indicated in Figure 5(a) and (b) that the measured SHG and SFG conversion efficiencies (red dots) increase linearly with the input pump power. Through the linear fitting (blue line), the normalized SHG and SFG conversion efficiencies are determined to be 0.46 %/mW and 0.48 %/mW, respectively.

In summary, we have demonstrated an on-chip Yb$^{3+}$-doped LN microdisk laser. We have measured the intrinsic quality factors of 3.56 × 10$^5$ at 976 nm and 1.1 × 10$^6$ at 1514 nm in the fabricated Yb$^{3+}$-doped LN microdisk. We observe the multi-mode laser emissions with a slope efficiency of 0.53% in the band from 1020 nm to 1070nm when pumped at 984 nm. The Yb$^{3+}$-doped LN microdisk lasers feature a low pump threshold of 103 μW at room temperature. Thanks to the strong χ$^{(2)}$ nonlinearity of LN, both intracavity SHG and SFG processes are observed in the Yb$^{3+}$-doped LN microdisk. The on-chip Yb$^{3+}$-doped LN microdisk laser offers great potential for biosensing application and integrated photonic chip.

**Funding**.
National Key R&D Program of China (2019YFA0705000), National Natural Science Foundation of China (Grant Nos. 12004116, 11874154, 11734009, 11933005, 11874060, 61991444), Shanghai Municipal Science and Technology Major Project (Grant No.2019SHZDZX01), Shanghai Sailing Program (21YF1410400), the Fundamental Research Funds for the Central Universities.

**Disclosures.** The authors declare no conflicts of interest